\documentclass[12pt,a4paper]{article}
\usepackage[utf8]{inputenc}

\usepackage{amsfonts,amssymb,amsmath,latexsym}
\usepackage{graphicx,graphics,epsfig}
\usepackage{hyperref}
\usepackage{xcolor}
\usepackage{float}
\usepackage{booktabs}
\usepackage{tabularx}
\usepackage{dcolumn}

\usepackage[T1]{fontenc}
\usepackage{tikz}
\usetikzlibrary{shapes.geometric, arrows}
\usetikzlibrary{arrows.meta, positioning}
\usepackage{comment}
\usepackage{setspace}
\usepackage{chngcntr}
\usepackage{setspace}
\usepackage{multirow}%
\usepackage{amsthm}%
\usepackage[title]{appendix}%
\usepackage{textcomp}%
\usepackage{authblk}

\usepackage{adjustbox}
\usepackage{threeparttable}
\usepackage{tikz}
\usetikzlibrary{positioning}
\usetikzlibrary{arrows.meta}

\hypersetup{
    colorlinks=true,
    linkcolor=blue,
    filecolor=magenta,
    urlcolor=blue,
}
\begin{document}

\title{Effects of interviewers on response to income and wealth items}
\author{
Moslem Rashidi\thanks{Department of Economics, University of Bologna, Piazza Scaravilli, 40126 Bologna, Italy (email: moslem.rashidi2@unibo.it).} \quad
\vspace{5mm}\\
}
\date{\today}

\maketitle

\begin{abstract}

Item nonresponse to financial questions is a persistent source of survey error, especially in interviewer-administered surveys. We examine whether interviewers’ expectations about respondents’ willingness to report income are associated with actual item responses to income and asset questions in Wave 6 of the Survey of Health, Ageing and Retirement in Europe (SHARE). Using data from 41,934 respondents in 12 countries, linked to interviewer survey and roster information, we analyze responses to four financial items with substantial nonresponse. We compare three approaches to handling missing covariates: complete-case analysis, multiple imputation (fill-in methods), and a generalized missing-indicator framework with information-criterion-based model averaging. Across most specifications, respondents interviewed by interviewers with higher expected income response rates are more likely to provide financial information. However, model averaging does not yield clear gains over simpler approaches. The results suggest that interviewer expectations contain useful information for understanding and modeling item nonresponse to sensitive financial items, with potential implications for interviewer training and survey fieldwork design.\end{abstract}

\begin{center}

{\bf Keywords}: SHARE; item nonresponse; missing data; logit; complete-case analysis; multiple imputation; model averaging

\bigskip

{\bf JEL classification}: To follow

\end{center}

\vfill

\thispagestyle{empty}
%\setcounter{page}{1}
%\baselineskip 19pt
%----------------------------------------------------------------------

%----------------------------------------------------------------------
\newpage
\section{Introduction}

Sample surveys frequently suffer from various sources of nonsampling error, such as coverage error, unit and item nonresponse, attrition, and measurement error, which may affect sample representativeness and data quality. These errors may depend on several features of the interview process (e.g., interview mode, length of the survey period, interviewers, interview instruments, question wording, and so on) (Groves and Couper 1998; West and Blom 2017; Banks \emph{et al.}\ 2011; Olson 2014).

In interviewer-administered surveys, interviewers play an important role in determining unit and item nonresponse outcomes (Korbmacher \emph{et al.}\ 2013; Friedel \emph{et al.}\ 2019; Durrant \emph{et al.}\ 2010; Tourangeau and Yan 2007; Pickery and Loosveldt 2001; Essig and Winter 2009). While some studies investigate whether interviewers’ sociodemographic characteristics, such as age, gender, race, ethnicity, education level, and experience, influence unit and item nonresponse (Berk and Bernstein 1988; Vercruyssen \emph{et al.}\ 2017; Riphahn and Serfling 2005; Bergmann \emph{et al.}\ 2022), there is substantial empirical evidence that non-demographic interviewer characteristics (e.g., personality traits, attitudes, and related factors) also affect unit and item nonresponse (see Lynn \emph{et al.}\ 2013; Blom and Korbmacher 2013; Lipps and Pollien 2011; Silber \emph{et al.}\ 2021; Wuyts and Loosveldt 2017; Schraepler 2006).

Information on interviewer characteristics is important for several reasons. First, it helps assess the potential for reducing nonsampling errors through targeted interviewer training, which may lower the occurrence of such errors (Groves and Couper 1998; Schaeffer \emph{et al.}\ 2010). Second, because interviewer characteristics are important determinants of response probabilities, they can be used in ex-post adjustment methods (e.g., weighting and imputation). In particular, they are relevant for adjustments based on the missing at random assumption. As emphasized by Fitzgerald \emph{et al.}\ (1998), Nicoletti and Peracchi (2005), and De Luca and Peracchi (2012), interviewer characteristics may also provide valid exclusion restrictions to identify more general missing data mechanisms.

In this paper, we use data from the sixth wave of the Survey of Health, Ageing and Retirement in Europe (SHARE) and the associated interviewer survey (SHARE\_IWS) to study how interviewers’ expectations about responses to income questions affect the actual response rates observed in the field. As shown by Sudman \emph{et al.}\ (1977) and Singer and Kohnke-Aguirre (1979), interviewers who anticipate difficulties tend to obtain lower response rates on sensitive questions, such as those related to gambling, income, alcohol consumption, mental health, and sexual behavior.

Friedel (2020) shows that interviewer expectations influence nonresponse rates to income and asset questions, and Cunha \emph{et al.}\ (2022) find that optimistic and self-confident interviewers achieve higher income response rates. Our paper is closely related to Friedel (2020), who studies the role of interviewer expectations in nonresponse to income and asset questions using data from the fifth wave of SHARE. A key difference is that we extend the analysis to the sixth wave of SHARE, in which the interviewer survey (IWS) covers 12 countries.

To study interviewer effects on nonresponse to income and wealth questions, we combine the interviewer survey data from the 2015 wave with an auxiliary data source, namely the interviewer roster database.\footnote{The interviewer roster data are not included in the public release of SHARE. This additional source of data was kindly provided by the SHARE central administration.} This dataset contains additional information on interviewer sociodemographic characteristics (e.g., age, gender, and years of experience). Although interviewers are not randomly assigned to respondents, interviewer characteristics are collected prior to the main survey, which helps mitigate concerns about reverse causality.

We also examine the role of cross-country heterogeneity in the relationship between interviewers’ expected response rates and actual response probabilities. To address missing covariate values, we compare three approaches: complete-case analysis (CCA), fill-in methods based on multiple imputation (FI), and the generalized missing-indicator (GMI) approach. We then apply the generalized missing-indicator framework proposed by Dardanoni \emph{et al.}\ (2011, 2012, 2015) to account for uncertainty arising from imputed values. In addition, we analyze missing data patterns arising from both interviewer nonparticipation and item nonresponse in the interviewer survey, as well as item nonresponse in the main SHARE interview. In particular, we extend the SHARE multiple imputation database for respondents’ characteristics to 100 imputations and supplement it with hot-deck multiple imputations for interviewer characteristics.

The remainder of the paper is organized as follows. Section~\ref{sec:SHARE} briefly describes the SHARE data for wave~6 and presents summary statistics of the response variables. Section~\ref{sec:predictors and missing data patterns}, titled ``Choice of predictors and missing data patterns,'' describes the regressors and their summary statistics. The statistical methods are presented in Section~\ref{sec:Methodology}. The empirical results are discussed in Section~\ref{sec:Results_2_}. Finally, Section~\ref{sec:CONCLUSIONS_2_} concludes.
%----------------------------------------------------------------------

%----------------------------------------------------------------------
\section{SHARE data}
\label{sec:SHARE}

This paper is based on release 7.1.0 of the Survey of Health, Ageing and Retirement in Europe (SHARE), a multidisciplinary and cross-national panel database of microdata on health, socioeconomic status, and social networks of the older European population. The panel currently comprises seven regular waves (2004--05, 2006--07, 2011, 2013, 2015, 2017, and 2019), which collect information on current living circumstances, and two retrospective waves (2008--09 and 2017), which focus on life histories.

We focus on the 2015 regular wave (i.e.\ wave~6), in which survey data on respondents are supplemented by auxiliary survey data on interviewers (the so-called interviewer survey, IWS) in two-thirds of the participating countries (Austria, Belgium, Estonia, Germany, Greece, Italy, Luxembourg, Poland, Portugal, Slovenia, Spain, and Sweden). To reduce missing values in basic sociodemographic characteristics of interviewers, we also combine the IWS data with additional paradata obtained from national survey agencies (the so-called interviewer roster, IWR). In the following subsections, we describe these two data sources and the criteria used to select our sample.

%----------------------------------------------------------------------

%----------------------------------------------------------------------
\subsection{Main survey data of SHARE wave 6}

The target population of wave~6 consists of individuals born in 1964 or earlier who speak one of the country's official languages (regardless of nationality or citizenship) and who do not reside abroad or in institutions such as prisons or hospitals during the entire fieldwork period.

National samples are selected using probability-based sampling designs. However, sampling procedures are not fully standardized across countries due to the lack of suitable sampling frames for the target population (see, e.g., Bergmann \emph{et al.}\ 2017). To limit the impact of representativeness issues and coverage errors for certain population groups, we restrict our sample to respondents born between 1934 and 1964 who live in residential households. Younger cohorts are included only because they are spouses or partners of age-eligible respondents and are therefore not representative of the underlying population. Similarly, we exclude older cohorts and individuals living in nursing homes or other healthcare institutions due to likely coverage errors in national sampling procedures for the institutionalized population. Overall, our sample includes 41,934 respondents from 12 countries that also participated in the SHARE interviewer survey for wave~6.

As in other regular waves of SHARE, the interview mode in wave~6 is face-to-face computer-assisted personal interviewing (CAPI), supplemented by show cards and a self-administered paper-and-pencil questionnaire. The CAPI questionnaire, which constitutes the largest part of the interview, is organized into 23 modules covering a wide range of topics, including demographics and family composition, physical and mental health, behavioral risks, cognitive abilities, well-being, labor force participation, income, health and consumption expenditures, assets, financial transfers, social relations, and expectations.

To reduce respondent burden, seven modules are administered to only one person per household or couple. Questions on assets and financial transfers are asked only to the financial respondent; questions on children and social support are asked only to the family respondent; and questions on the income of non-eligible household members, housing, and consumption expenditure are asked only to the household respondent.\footnote{The ``financial respondent'' is either a single individual or the partner within a couple who is most knowledgeable about financial matters. The ``family respondent'' is either a single individual or the partner within a couple who is interviewed first, while the ``household respondent'' is the household member most knowledgeable about housing matters.} An additional exception is the module on interviewer observations, which collects information on the interview process and is completed by the interviewer at the end of each interview without involving the respondent.

Most variables in SHARE exhibit negligible fractions of missing values (typically well below 5 percent of eligible respondents for each item). However, this is not the case for financial variables related to income and assets, which are collected through open-ended questions that are sensitive and difficult to answer. Table~\ref{tab:SUMMARY_STAT_Y} reports response rates for four financial variables in the sample of eligible respondents, including income from various sources (first panel) and real and financial assets (second panel). The number of eligible respondents, after excluding outliers, varies across items due to branching and skip patterns in the CAPI questionnaire and is obtained from the imputations module (gv\_imputations).

Response rates for some financial variables are particularly low and therefore concerning. For example, approximately one-third of eligible respondents do not answer questions about money held in bank accounts or the value of their main residence. In addition, a single question on total household income exhibits about 24 percent missing values. These substantial levels of missingness raise serious concerns about potential selection bias and efficiency losses arising from item nonresponse.

\begin{table}[ht]
\caption{Response rate of answers on the financial variables of SHARE wave 6 in the eligible respondent's sample}
\begin{center}
\begin{tabular}{lrrrrr}
\hline
\multicolumn{1}{l}{ }&
\multicolumn{3}{c}{Respondents}&
\multicolumn{1}{c}{}&
\multicolumn{1}{c}{}\\
\cline{2-4}
\multicolumn{1}{l}{Variable}&
\multicolumn{1}{c}{Type}&
\multicolumn{1}{c}{Elig.}&
\multicolumn{1}{c}{RR}\\
\hline
Total household income & HR                          &             28204 &        0.761 \\
Old age, early ret., survivor pensions  &AR         &             23615 &        0.848 \\
\hline
Bank accounts   & HR                                 &             24919 &        0.615 \\
Value of main residence & HR                         &             22341 &        0.660 \\
\hline
\end{tabular}
\parbox{146mm}{\footnotesize {\em Notes}.
AR means ``all respondents'', HR means ``household respondents''.
RR is the response rate on each financial variable.}
\end{center}
\label{tab:SUMMARY_STAT_Y}
\end{table}

In this paper, we study the determinants of the response process for the four financial variables listed in Table~\ref{tab:SUMMARY_STAT_Y}. In addition to the observable characteristics of eligible respondents for each item, we exploit auxiliary data collected in the SHARE interviewer survey to evaluate the impact of observable interviewer characteristics on the probability of responding to financial variables.

%----------------------------------------------------------------------

%----------------------------------------------------------------------
\subsection{Interviewer survey data of SHARE wave 6}

\begin{table}[th]
\caption{Number of interviewers, number of participants, and participation rate to the interviewer survey (IWS) of SHARE wave 6 by country}
\begin{center}
\begin{tabular}{lrrrrr}
\hline
\multicolumn{1}{l}{} &
\multicolumn{1}{c}{}   &
\multicolumn{2}{c}{IWS} &
\multicolumn{2}{c}{}\\
\cline{3-4}
\multicolumn{1}{l}{Country} &
\multicolumn{1}{c}{Total}   &
\multicolumn{1}{c}{$\,$Obs.$\,$}     &
\multicolumn{1}{c}{$\,$PR$\,$}  \\
\hline
       Austria     &       70     &         51     &    0.729     \\  
       Belgium     &      132     &        106     &    0.803     \\  
       Estonia     &       82     &         35     &    0.427     \\  
       Germany     &      147     &        128     &    0.871      \\  
        Greece     &      170     &         88     &    0.518      \\  
         Italy     &      140     &        132     &    0.943     \\  
    Luxembourg     &       44     &         24     &    0.545     \\  
        Poland     &       60     &         27     &    0.450      \\  
      Portugal     &       51     &         39     &    0.765     \\  
      Slovenia     &       59     &         48     &    0.814      \\  
         Spain     &      116     &         57     &    0.491      \\  
        Sweden     &      101     &         73     &    0.723      \\  
 \hline
       Total     &     1172     &        808     &    0.689      \\  
 \hline
\end{tabular}
\parbox{94mm}{\footnotesize {\em Notes}.
PR denotes the participation rate to the IWS.}
\end{center}
\label{tab:IW_PART}
\end{table}

Since its pilot implementation in wave~4, SHARE has conducted an interviewer survey (IWS) to supplement respondent survey data with detailed information on interviewers. The IWS for wave~6 was administered as an online survey after national interviewer training sessions but prior to the fieldwork of the main survey in each country. Although interviewers are not randomly assigned to respondents, this feature of the IWS ensures that the variables used to study interviewer effects on survey outcomes are not subject to reverse causality.

In addition to basic sociodemographic characteristics such as gender, age, educational attainment, and occupational status, the IWS questionnaire is based on the conceptual framework developed by Blom and Korbmacher (2013), which identifies four key dimensions of interviewer characteristics relevant for understanding interviewer effects on various forms of nonsampling error, including unit and item nonresponse, lack of consent to record linkage, and lack of cooperation with other survey requests. In this study, we focus on interviewer characteristics that are likely to play an important role in explaining item nonresponse to financial questions.

The first dimension of the IWS questionnaire captures interviewers' attitudes toward the survey process and their job. These are measured through questions on the reasons for becoming an interviewer, the circumstances under which interviewers deviate from the interview protocol to better approach difficult respondents, trust in others, and data protection concerns. The second dimension concerns interviewers’ own behavior regarding data collection requests and how they would behave in situations similar to those faced by respondents. For example, the IWS includes questions on whether interviewers should respect respondents' privacy, whether refusals from reluctant respondents should be accepted, and whether exerting substantial effort to persuade respondents affects the reliability of their answers. In addition, interviewers are asked to report their total household income to assess potential relationships between the response behavior of interviewers and respondents on sensitive financial questions. The third dimension captures interviewers' experience with social surveys in general and with previous waves of SHARE. Finally, the fourth dimension relates to interviewers’ expectations about survey outcomes, such as their expectations of response rates to income questions.

The IWS provides valuable information for understanding the complex process through which interviewers may influence nonsampling errors. However, because the IWS itself is subject to unit and item nonresponse, estimates of interviewer effects may be biased and inefficient. In SHARE, participation in the IWS is voluntary at both the country and interviewer levels. Table~\ref{tab:IW_PART} reports the number of interviewers involved in the main survey for wave~6, the number who participated in the IWS, and the corresponding participation rates by country. Overall, the IWS includes 808 of the 1,172 interviewers who conducted at least one interview in the main survey. The average participation rate across countries is 69 percent, ranging from 43 percent in Estonia to 94 percent in Italy.

To mitigate the impact of nonresponse in the IWS, we also exploit interviewer roster (IWR) data collected by national survey agencies. These administrative data contain limited information on interviewer characteristics, including gender, age, years of education, years of experience, and participation in previous waves of SHARE. However, they cover a larger number of interviewers and thus provide additional information on missing interviewer characteristics. IWR data are not available for all Swedish interviewers, as well as for 17 interviewers in Greece and 18 in Portugal. Moreover, some interviewer characteristics are not observed in all countries (e.g., years of education is unavailable in Germany, and years of experience is unavailable in Poland).
%----------------------------------------------------------------------

%----------------------------------------------------------------------
\section{Choice of predictors and missing data patterns}
\label{sec:predictors and missing data patterns}
In this study, we model the probability of responding to financial questions in surveys as a function of interviewers' expectations regarding response rates to income questions, along with a set of control variables.

\begin{figure}[htbp]
	\centering
		\includegraphics[width=4.9in]{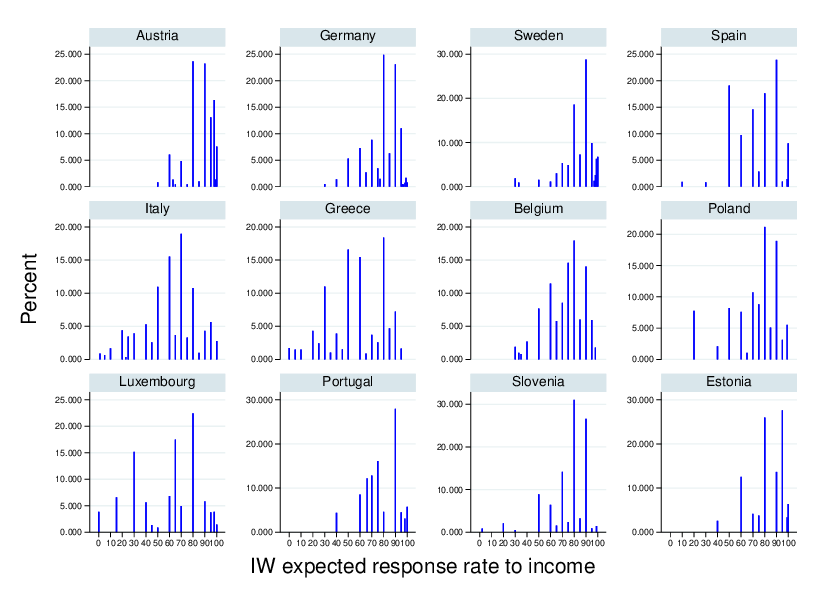}
	\caption{Distribution of expectations of the interviewer on response rate to income by country}
	\label{Fig:IW_ERRI}
\end{figure}

{\bf Regressor of interest.} The main explanatory variable in our study is the interviewer's expectation of the probability that respondents provide meaningful answers to financial questions. Before the start of the fieldwork period, interviewers are asked the following question: ``What does the interviewer expect? How many of his/her respondents (in percentage) in SHARE will provide information about their income?'' Interviewers are asked to provide a numerical response ranging from 0 to 100 percent, in one percentage-point increments. The empirical distribution of this variable exhibits several focal values (see Figure~\ref{Fig:IW_ERRI}). To mitigate the impact of measurement error, we construct a binary indicator equal to 1 for interviewers whose expected response rate exceeds the median of the country-specific distribution.

{\bf Control variables.} In addition to basic sociodemographic characteristics (e.g., gender, age, years of experience, and education level), interviewer-level controls include workload status during the fieldwork period, self-reported health status, and the interviewer's response to the total household income (THI) question, indicating whether the interviewer reports his/her own household income after taxes. We also include variables capturing interviewer behavior during the CAPI interview, such as whether the interviewer tends to speak quickly when the respondent appears to be in a hurry and whether the interviewer clarifies questions when respondents do not understand them.

At the respondent level, control variables include sociodemographic characteristics such as gender, age, marital status, education level, and participation in previous waves. Additional covariates capturing respondent characteristics include measures of cognitive ability (numeracy, fluency, and self-rated memory) and indicators of physical and mental health (self-assessed health status, body mass index, and depression status). Definitions and summary statistics for the main explanatory variable and control variables are reported in Table~\ref{tab:SUMMARY_STAT_IW}.

\begin{table}[ht]
\caption{Definitions and summary statistics of the control variables in the eligible respondents' sample}
\begin{center}
\begin{tabular}{l|rrrr}
\hline
\multicolumn{1}{l}{Description}&
\multicolumn{1}{c}{Obs.}&
\multicolumn{1}{c}{Mean}&
\multicolumn{1}{c}{Std.}\\
\hline
IW ERR income: high                             & 26577 & 0.439 & 0.496\\
IW female                                       & 40537 & 0.721 & 0.449\\
IW high education                               & 29213 & 0.432 & 0.495\\
IW workload: high                               & 41934 & 0.749 & 0.434\\
IW good health                                  & 29386 & 0.585 & 0.493\\
IW response to THI                              & 29498 & 0.702 & 0.458\\
IW speak fast                                   & 29397 & 0.425 & 0.494\\
IW clarifies questions                          & 29449 & 0.607 & 0.489\\
IW age                                          & 40537 & 51.388 & 11.767\\
IW yrs of experience                            & 38296 & 10.314 & 8.895\\
R female                                        & 41934 & 0.552 & 0.497\\
R lives in couple                               & 41934 & 0.758 & 0.428\\
R high education                                & 41228 & 0.592 & 0.491\\
R numeracy score                                & 38128 & 0.657 & 0.475\\
R good health                                   & 41870 & 0.613 & 0.487\\
R part. past waves                              & 41934 & 0.807 & 0.395\\
R good memory                                   & 39006 & 0.731 & 0.444\\
R limited activities                            & 41868 & 0.443 & 0.497\\
R depression status                             & 40230 & 0.730 & 0.444\\
R age                                           & 41929 & 65.806 & 8.120\\
R fluency score                                 & 40216 & 20.008 & 7.860\\
R BMI                                           & 41112 & 27.146 & 4.594\\
\hline
\end{tabular}
\parbox{146mm}{\footnotesize {\em Obs.\ denotes the number of observations in the eligible respondents' sample, Mean denotes the average, and Std.\ denotes standard deviations.}}
\end{center}
\label{tab:SUMMARY_STAT_IW}
\end{table}

{\bf Missing data patterns.} In our study, most covariates exhibit missing values arising from two main sources:

\begin{enumerate}

\item {\bf Nonresponse in the interviewer survey (IWS).} Missing data arise from interviewer nonparticipation or item nonresponse within the IWS. This source of missingness leads to a substantial proportion of missing data at the respondent level. Let $\text{I}_1$ be a binary indicator equal to 1 for interviewers who participate in the IWS and provide complete responses, and 0 otherwise.

\item {\bf Nonresponse in the CAPI survey.} This source concerns variables related to respondents' sociodemographic characteristics, health measures, and cognitive abilities. Let $\text{I}_2$ be a binary indicator equal to 1 for respondents who participate in the CAPI survey and provide complete responses, and 0 otherwise.

\end{enumerate}

Given these two sources of missing data, three distinct missing data patterns arise across countries, as reported in Table~\ref{TAB:MDPATT_CC}.
%----------------------------------------------------------------------

%----------------------------------------------------------------------
\section{Methodology}
\label{sec:Methodology}
In this section, we examine three approaches for handling missing covariate values.

\subsection{Complete-case analysis}

One common approach to handling missing data is complete-case analysis (CCA), which consists of deleting all observations with missing covariate values and estimating the model of interest using only the complete-case (CC) subsample. In our application, this corresponds to estimating a set of logit models for the binary response indicators of the financial outcomes listed in Table~\ref{tab:SUMMARY_STAT_Y}. Let $\text{Y}_{j,0}$ $(\text{N}_{j,0}\times 1)$ denote the vector of observations on the outcome of interest in the complete-case subsample. This variable takes the value 1 if the $i$th eligible respondent answers the $j$th financial question, and 0 otherwise. The logistic regression model for the complete-case subsample can then be written as the following linear predictor:
\begin{equation}
\eta_{j,0}=\textbf{X}_{j,0}^\top\beta_j,
\label{eq:CCA_model}
\end{equation}
where $\textbf{X}_{j,0}$ $(\text{N}_{j,0}\times \text{K}_{j})$ is the matrix of regressors observed in the complete-case subsample, with $\text{N}_{j,0}<\text{N}_j$, and $\beta_j$ $(\text{K}_{j}\times 1)$ is the unknown parameter vector.

In addition to the standard regularity conditions required for maximum likelihood (ML) estimation of binary logit models, the properties of the CCA estimator of $\beta_j$ in model~\eqref{eq:CCA_model} depend crucially on the validity of two assumptions (Dardanoni \emph{et al.}\ 2015): 
(i) the Fisher information matrix for the complete-case subsample is positive definite with probability approaching one as the sample size tends to infinity; and 
(ii) conditional on the covariates, the response probability is the same in the subsamples with and without missing covariates. The first assumption ensures that $\beta_j$ is identified from the complete-case subsample for each outcome. The second requires the response variable and the missing-data mechanism for the covariates to be conditionally independent given the observed covariates.

Even when these two assumptions hold and CCA yields asymptotically consistent estimates of $\beta_j$, it discards a substantial amount of data. In our study, the complete-case subsample contains 22,609 observations, as shown in Table~\ref{TAB:MDPATT_CC}, whereas the full sample contains 41,934 observations. Thus, using CCA implies the loss of 19,325 observations. This loss of information reduces precision. Hence, although CCA is simple to implement, it comes at the cost of discarding potentially valuable information, which motivates considering imputation-based approaches.

\subsection{Fill-in approach}

The fill-in (FI) approach is a popular way to handle missing data by replacing missing values with estimated ones. This approach includes several methods, but here we focus on the most widely used one, namely multiple imputation (MI). Let $\text{Y}_j$ $(\text{N}_j\times 1)$ denote the vector of observations on the outcome of interest in the full sample. The logit model estimated on the fill-in sample can be written as the following linear predictor:
\begin{equation}
\eta_j=\textbf{W}_j^\top\beta_j,
\label{eq:FI_model}
\end{equation}
where $\textbf{W}_j$ $(\text{N}_j\times \text{K}_j)$ is the matrix of observed and imputed regressors, and $\beta_j$ $(\text{K}_j\times 1)$ is the unknown parameter vector.

Under multiple imputation, obtaining an asymptotically equivalent fill-in ML estimator to the ML estimator from the complete-data sample requires, in addition to the missing-at-random (MAR) assumption, that the imputation model be more general than the analysis model, that is, that the two models be congenial (Meng 1994). However, the validity of the imputation model cannot be taken for granted. Accordingly, the FI parameter estimates, denoted by $\beta_{FI\_MI}$, may fail to be asymptotically consistent.

In our application, missing covariate values for respondents' and interviewers' characteristics are imputed using the fully conditional specification (FCS) method of van Buuren \emph{et al.}\ (2006), which is based on an iterative sequence of univariate imputation models. To ensure that the imputed values for a given interviewer remain the same across all respondents assigned to that interviewer, we use two sequential Gibbs samplers: one for respondents' variables and one for interviewers' variables. In addition, standard errors may decrease either because of a stronger association between the auxiliary variables used in the imputation model and the variables being imputed or because of a larger number of imputations (von Hippel 2020). To choose a minimally sufficient number of imputations, we use the rule of thumb $\text{M} \geq 100\times \text{FMI}$, where FMI denotes the fraction of missing information, that is, the share of information lost because of missing data (see, e.g., the Stata 17 help manual). Since the FMI for interviewer variables is around 0.85, we use $\text{M}=100$ imputations.

\subsection{Generalized missing-indicator (GMI) and model averaging (MA)}

The generalized missing-indicator approach, or ``grand model,''\footnote{We consider the grand model of the form
\begin{equation}
\eta_j=\textbf{W}_j^\top\beta_j+\textbf{Z}_j^\top\delta_j,
\label{eq:Grand_model}
\end{equation}
where $\textbf{W}_j$ $(\text{N}_j\times \text{K}_j)$ and $\textbf{Z}_j$ $(\text{N}_j\times \text{H}_j\text{K}_j)$ are the matrices of ``fill-in'' and ``auxiliary'' regressors, respectively.} introduced by Dardanoni \emph{et al.}\ (2011, 2012, 2015) enlarges the model space by considering not only the unrestricted and fully restricted versions of the grand model, corresponding respectively to the fill-in (FI) approach in~\eqref{eq:FI_model} and the complete-case (CC) approach in~\eqref{eq:CCA_model},\footnote{Dardanoni \emph{et al.}\ (2015) show that the complete-case analysis estimate of $\beta_j$ in model~\eqref{eq:CCA_model} is numerically equivalent to the ML estimate of $\beta_j$ in the grand model.} but also all intermediate submodels in which a subset of the auxiliary parameters $\delta_j$ is constrained to zero. The key implication of this enlarged model space is the emergence of model uncertainty. One way to address model uncertainty is model averaging (MA), based on the idea that each model contributes information about the parameters of interest.

Let all possible missing-data patterns be indexed by $\text{h}_j=\{1,\cdots,\text{H}_j\}$. The model space $\mathcal{M}_j$ then includes $\text{R}_j$ possible logit models, where $\text{R}_j=2^{\text{H}_j}$, that is, $\mathcal{M}_j=\{\text{M}_{j,1},\cdots,\text{M}_{j,\text{R}_j}\}$. The $r_j$th logit model, $\text{M}_{j,r_j}$, can be written as
\begin{equation}
\eta_{j,r_j}=\textbf{W}_j^\top\beta_j+\textbf{Z}_{j,r_j}^\top\delta_{j,r_j},
\label{eq:G_models}
\end{equation}
where $\textbf{W}_j$ $(\text{N}_j\times \text{K}_j)$ is the matrix of ``focus'' regressors and $\textbf{Z}_{j,r_j}$ is the matrix collecting the subset of $\mathsf{P}_{j,r_j}\in[0,\text{H}_j\text{K}_j]$ auxiliary regressors included in model $r_j$. The vector $\delta_{j,r_j}$ contains the corresponding auxiliary coefficients.

The model-averaging estimator of the coefficient of interest, $\beta_{j,\text{MA}}$, is given by
\begin{equation}
\widehat{\beta}_{j,\text{MA}}=\sum_{r_j=1}^{R_j}\lambda_{j,r_j}\widehat{\beta}_{j,r_j},
\label{eq:BMA_model1}
\end{equation}
where $\lambda_{j,r_j}\geq 0$ and $\sum_{r_j=1}^{R_j}\lambda_{j,r_j}=1$. Here, $\widehat{\beta}_{j,r_j}$ denotes the ML estimator of $\beta_j$ under model $r_j$.

\subsubsection{Bayesian model averaging (BMA) with information-criterion weights}

In this study, we consider two common approaches for approximating the posterior model probabilities, denoted by $\lambda_{j,r_j}=\Pr(\text{M}_{j,r_j}\mid \text{Y}_j)$. The weights used in the model-averaging estimator of $\beta_{j,\text{MA}}$ are
\begin{equation}
\lambda_{j,r_j}=\Pr(\text{M}_{j,r_j}\mid \text{Y}_j)=
\frac{\Pr(\text{Y}_j\mid \text{M}_{j,r_j})\Pr(\text{M}_{j,r_j})}
{\sum_{r_j=1}^{R_j}\Pr(\text{Y}_j\mid \text{M}_{j,r_j})\Pr(\text{M}_{j,r_j})},
\label{eq:P_M}
\end{equation}
where $\Pr(\text{M}_{j,r_j})$ is the prior probability of model $r_j$, and
\begin{equation}
\Pr(\text{Y}_j\mid \text{M}_{j,r_j})=
\int_{\beta_j\in \mathbf{B}_j}\Pr(\text{Y}_j\mid \beta_{j,r_j},\text{M}_{j,r_j})
\Pr(\beta_{j,r_j}\mid \text{M}_{j,r_j})\,d\beta_{j,r_j},
\label{eq:marginalLiklihood_model}
\end{equation}
is the marginal likelihood of model $r_j$. In this expression, $\Pr(\text{Y}_j\mid \beta_{j,r_j},\text{M}_{j,r_j})$ is the sample likelihood, $\Pr(\beta_{j,r_j}\mid \text{M}_{j,r_j})$ is the prior density of $\beta_{j,r_j}$ under model $r_j$, and $\beta_{j,r_j}$ is the vector of model-specific parameters of interest.

To compute the posterior model probability in~\eqref{eq:P_M}, one must evaluate the marginal likelihood in~\eqref{eq:marginalLiklihood_model}. In general, however, no closed-form solution is available. To address this problem, Raftery (1996) uses a Laplace approximation. Under diffuse priors and equal prior model probabilities, Schwarz's theorem (1978) implies that the marginal likelihood of model $r_j$ can be approximated as
\[
\Pr(\text{Y}_j\mid \text{M}_{j,r_j})\approx \exp\left(-\frac{\text{BIC}_{j,r_j}}{2}\right).
\]
The BMA weights can then be approximated by
\begin{equation}
\lambda_{j,r_j}=\Pr(\text{M}_{j,r_j}\mid \text{Y}_{j})=
\frac{\exp(-\text{BIC}_{j,r_j}/2)}
{\sum_{r_j=1}^{R_j}\exp(-\text{BIC}_{j,r_j}/2)}
=
\frac{\exp(-\Delta \text{BIC}_{j,r_j}/2)}
{\sum_{r_j=1}^{R_j}\exp(-\Delta \text{BIC}_{j,r_j}/2)},
\label{eq:posteriorModel_model1}
\end{equation}
where
\begin{equation}
\Delta \text{BIC}_{j,r_j}=\text{BIC}_{j,r_j}-\text{BIC}_{j,r_j,\min},
\end{equation}
and $\text{BIC}_{j,r_j,\min}$ is the minimum BIC value across the $R_j$ competing models.

In the spirit of likelihood-ratio methods, Buckland \emph{et al.}\ (1997) propose an alternative approximation to posterior model probabilities based on Akaike's Information Criterion (AIC). The corresponding weights are
\begin{equation}
\lambda_{j,r_j}=\Pr(\text{M}_{j,r_j}\mid \text{Y}_j)=
\frac{\exp(-\text{AIC}_{j,r_j}/2)}
{\sum_{r_j=1}^{R_j}\exp(-\text{AIC}_{j,r_j}/2)}.
\label{eq:posteriorModel_model2}
\end{equation}
A larger value of $\lambda_{j,r_j}$ indicates a more plausible model. For ease of computation, Burnham and Anderson (2002) propose the equivalent formulation
\begin{equation}
\lambda_{j,r_j}=\Pr(\text{M}_{j,r_j}\mid \text{Y}_j)=
\frac{\exp(-\Delta \text{AIC}_{j,r_j}/2)}
{\sum_{r_j=1}^{R_j}\exp(-\Delta \text{AIC}_{j,r_j}/2)},
\label{eq:posteriorModel_model3}
\end{equation}
where
\begin{equation}
\Delta \text{AIC}_{j,r_j}=\text{AIC}_{j,r_j}-\text{AIC}_{j,r_j,\min},
\end{equation}
and $\text{AIC}_{j,r_j,\min}$ is the minimum AIC value across the $R_j$ competing models.

%----------------------------------------------------------------------
\section{Results}
\label{sec:Results_2_}

Overall, the findings show that, for all four financial questions considered---household income, old-age/early-retirement/survivor pensions, bank accounts, and the value of the main residence---higher interviewer expected response rates are generally associated with lower item nonresponse across the approaches considered. The results are reported in Tables~\ref{TAB:MDPATT_CC}, \ref{TAB:AME_ERR_FI_MI}, \ref{TAB:AME_ERR_BBMA_BIC}, and \ref{TAB:AME_ERR_BBMA_AIC}. In only a small number of country-specific cases does the interviewer's expected response rate have a negative effect on response behavior, and statistically significant negative effects are rare. Significant negative effects appear only for the value of the main residence in Greece under the FI\_MI and BBMA\_BIC approaches, and for pensions in Spain under the CCA and BBMA\_AIC approaches.

A key motivation for using the fill-in approach is that it incorporates all available information into the logistic regression model, which may improve precision. In the multiple-imputation (FI\_MI) approach, however, the pooled standard error consists of two components: within-imputation variance and between-imputation variance. Although the squared standard errors within each imputed dataset, that is, the within-imputation variance, are smaller under FI\_MI than under CCA, the pooled standard errors under FI\_MI are often larger than those under CCA because they also reflect additional uncertainty arising from the imputation of missing values, namely the between-imputation variance (see Tables~\ref{TAB:MDPATT_CC} and \ref{TAB:AME_ERR_FI_MI}).

Finally, we compare the predictive performance of block Bayesian model averaging based on BIC and AIC weights with that of the CCA and FI\_MI estimators. Although all models in our model space are assumed to be equally likely \emph{a priori}, the posterior distribution produced by the block BMA procedures is concentrated on only a few models. As a result, the block BMA\_AIC estimates, $\beta_{BMA\_AIC}$, are approximately equal to the CCA estimates, $\beta_{CCA}$, while the block BMA\_BIC estimates, $\beta_{BMA\_BIC}$, are approximately equal to the FI\_MI estimates, $\beta_{FI\_MI}$.

{\bf Limitations.} Our results are subject to some limitations. First, the average marginal effects (AMEs) produced by block Bayesian model averaging based on BIC and AIC are very close to the two extreme specifications of the GMI grand model, namely CCA and FI\_MI. We therefore explored the use of conjugate priors for logit models, which would allow posterior model probabilities to be estimated using a Markov Chain Monte Carlo algorithm. However, the large number of imputed datasets makes this approach computationally costly. Future work may consider alternative approaches, such as WALS (Magnus and De Luca 2016), to reduce computation time substantially. Second, because respondents are nested within interviewers, a standard logit model is likely to underestimate the true standard errors, which may lead to over-rejection of the null hypothesis.
%----------------------------------------------------------------------

%----------------------------------------------------------------------
\section{Conclusions}
\label{sec:CONCLUSIONS_2_}
%----------------------------------------------------------------------
Our results support the hypothesis that interviewer expectations regarding respondents' willingness to report their income are associated with item nonresponse to financial questions. Respondents are more likely to provide information on income and assets when they are interviewed by an interviewer who expects more than 50 percent of respondents to report their income than when they are interviewed by an interviewer who expects 50 percent or fewer respondents to do so.

Moreover, in the present study, interviewers with more optimistic expectations are associated with reductions in item nonresponse of up to 14 percent for income questions in some countries and up to 26 percent for asset questions in some countries. These effects are substantial relative to the average response rates of 80 percent for the two income questions and 64 percent for the two asset questions considered. Overall, these findings indicate that interviewer expectations are important in surveys of older individuals. To reduce nonresponse to financial questions, our results suggest the importance of interviewer selection and targeted training strategies.
\clearpage
%----------------------------------------------------------------------

%-----------------------------------------------------------------------------------------

\section*{Acknowledgements}

This paper uses data from SHARE Wave~6 (DOI: 10.6103/SHARE.w6.700); see Börsch-Supan \emph{et al.}\ (2013) for methodological details. The SHARE data collection has been funded by the European Commission, DG RTD through FP5 (QLK6-CT-2001-00360), FP6 (SHARE-I3: RII-CT-2006-062193; COMPARE: CIT5-CT-2005-028857; SHARELIFE: CIT4-CT-2006-028812), FP7 (SHARE-PREP: GA N°211909; SHARE-LEAP: GA N°227822; SHARE M4: GA N°261982; DASISH: GA N°283646), and Horizon 2020 (SHARE-DEV3: GA N°676536; SHARE-COHESION: GA N°870628; SERISS: GA N°654221; SSHOC: GA N°823782; SHARE-COVID19: GA N°101015924), as well as by DG Employment, Social Affairs \& Inclusion through VS 2015/0195, VS 2016/0135, VS 2018/0285, VS 2019/0332, VS 2020/0313, SHARE-EUCOV: GA N°101052589, and EUCOVII: GA N°101102412. Additional funding from the German Federal Ministry of Research, Technology and Space (01UW1301, 01UW1801, 01UW2202), the Max Planck Society for the Advancement of Science, the U.S. National Institute on Aging (U01\_AG09740-13S2, P01\_AG005842, P01\_AG08291, P30\_AG12815, R21\_AG025169, Y1-AG-4553-01, IAG\_BSR06-11, OGHA\_04-064, BSR12-04, R01\_AG052527-02, R01\_AG056329-02, R01\_AG063944, HHSN271201300071C, RAG052527A), and various national funding sources are gratefully acknowledged (see \url{www.share-eric.eu}).)
%--------------------------------------------------------------------------------------------------------------------------------------------------------

%----------------------------------------------------------------------
%References
%----------------------------------------------------------------------

\newpage
\section*{References}
\begin{description}

\item Anderson, D.R. and Burnham, K.P. (2002).
Avoiding pitfalls when using information-theoretic methods.
\emph{The Journal of wildlife management} 66, 912--918.

\item Banks, J., Muriel, A., and Smith, J. (2011).
Attrition and health in ageing studies: Evidence from ELSA and HRS.
\emph{Longitudinal and Life Course Studies}, 2: 1--29.

\item Bergmann, M., Franzese, F., and Schrank, F. (2022). 
Determinants of consent in the SHARE accelerometer study.
\emph{SHARE Working Paper Series}, N. 78.

\item Bergmann, M., Kneip, T., De Luca, G. and Scherpenzeel, A., (2017). 
Survey participation in the survey of health, ageing and retirement in Europe (SHARE), Wave 1-6.
\emph{Munich: Munich Center for the Economics of Aging}.

\item Berk, M.~L., and Bernstein, A.~B. (1988).
Interviewer characteristics and performance on a complex health survey.
\emph{Social Science Research}, 17: 239--251.

\item Blom, A.~G., and Korbmacher, J.~M. (2013). 
Measuring interviewer effects in SHARE Germany.
\emph{SHARE Working Paper Series}, N. 3.

\item Buckland, S.T., Burnham, K.P. and Augustin, N.H. (1997). 
Model selection: an integral part of inference.
\emph{Biometrics} 53, 603--618.

\item Cunha, C., Matos, A.D., Voss, G. and Machado, C., 2022. Interviewer characteristics and nonresponse survey outcomes: A Portuguese case study.
\emph{In Challenges and Trends in Organizational Management and Industry. Springer}, Cham. 95--111

\item Dardanoni, V., Modica, S., and Peracchi, F. (2011).
Regression with imputed covariates: A generalized missing-indicator approach.
\emph{Journal of Econometrics}, 162: 362--368.

\item Dardanoni, V., De Luca, G., Modica, S. and Peracchi, F. (2012).
A generalized missing-indicator approach to regression with imputed covariates. \emph{The Stata Journal}, 12: 575-604.

\item Dardanoni, V., De Luca, G., Modica, S., and Peracchi, F. (2015).
Model averaging estimation of generalized linear models with imputed covariates.
\emph{Journal of Econometrics}, 184: 452--463.

\item De Luca, G., and Peracchi, F. (2012).
Estimating Engel curves under unit and item nonresponse.
\emph{Journal of Applied Econometrics}, 27: 1076--1099.

\item  Durrant, G.~B., Groves, R.~M., Staetsky, L., and Steele, F. (2010).
Effects of interviewer attitudes and behaviors on refusal in household surveys.
\emph{Public Opinion Quarterly}, 74: 1--36.

\item Essig, L. and Winter, J.K. (2009). 
Item non‐response to financial questions in household surveys: An experimental study of interviewer and mode effects.
\emph{Fiscal Studies}, 30: 367--390.

\item Fitzgerald, J., Gottschalk, P., and Moffitt, R. (1998).
An analysis of sample attrition in panel data: The Michigan panel study of income dynamics.
\emph{The Journal of Human Resources}, 33: 251--299

\item Friedel, S.\ ( 2020).
What they expect is what you get: The role of interviewer expectations in nonresponse to income and asset questions.
\emph{Journal of Survey Statistics and Methodology}, 8: 851--876.

\item  Friedel, S., Bethmann, A., and Kronenberg, M. (2019).
The third round of the SHARE interviewer survey.
\emph{SHARE Wave 7 Methodology: Panel Innovations and Life Histories}, 101--106.

\item Groves, R.~M., Couper, M.~P. (1998).
Nonresponse in household interview surveys.
\emph{John Wiley and Sons}.

\item Korbmacher, J.~M., Friedel, S., Wagner, M., and Krieger, U. (2013). Interviewing interviewers: The SHARE interviewer survey. 
\emph{SHARE Wave 5: Innovations \& Methodology}, 67--74.

\item Lipps, O., and Pollien, A. (2011).
Effects of interviewer experience on components of nonresponse in the European Social Survey. 
\emph{Field Methods}, 23: 156--172.

\item Lynn, P., Sinibaldi, J. and Tipping, S. (2013).
The effect of interviewer experience, attitudes, personality and skills on respondent co-operation with face-to-face surveys. 
\emph{Survey Research Methods}, 7: 1--15.

\item Magnus, J.~R., and De Luca, G. (2016).
Weighted-average least squares (WALS): A survey.
\emph{Journal of Economic Surveys} 30: 117--148.

\item Meng, X.L. (1994).
Multiple-imputation inferences with uncongenial sources of input. \emph{Statistical Science} 9, 538--558.

\item Nicoletti, C., and Peracchi, F. (2005). 
Survey response and survey characteristics: microlevel evidence from the European Community Household Panel.
\emph{Journal of the Royal Statistical Society}, 168: 763--781.

\item Olson, K. (2014).
Do non-response follow-ups improve or reduce data quality? A review of the existing literature.
\emph{Quality Control and Applied Statistics}, 59: 61--62.

\item Pickery, J., and Loosveldt, G. (2001). 
An exploration of question characteristics that mediate interviewer effects on item nonresponse. 
\emph{Journal of Official Statistics}, 17: 337--350

\item Raftery, A.E. (1996). 
Approximate Bayes factors and accounting for model uncertainty in generalised linear models.
\emph{Biometrika}, 83, 251--266.

\item Riphahn, R.T., and Serfling, O. (2005).
Item non-response on income and wealth questions.
\emph{Empirical Economics}, 30: 521--538.

\item Schaeffer, N.~C., Dykema, J., and Maynard, D.~W. (2010).
Interviewers and interviewing.
\emph{Handbook of Survey Research}, 2: 437--471.

\item Schraepler, J.~P. (2006). 
Explaining income nonresponse-a case study by means of the British Household Panel Study (BHPS). 
\emph{Quality and Quantity}, 40: 1013-1036.

\item Silber, H., Roßmann, J., Gummer, T., Zins, S. and Weyandt, K.W. (2021).
The effects of question, respondent and interviewer characteristics on two types of item nonresponse.
\emph{Journal of the Royal Statistical Society: Series A (Statistics in Society)}, 184: 1052--1069.

\item Singer, E. and Kohnke-Aguirre, L. (1979). 
Interviewer expectation effects: A replication and extension.
\emph{Public Opinion Quarterly}, 43: 245--260.

\item Sudman, S., Bradburn, N.M., Blair, E.D., and Stocking, C.\ (1977). 
Modest expectations: The effects of interviewers' prior expectations on responses.
\emph{Sociological Methods \& Research}, 6: 171--182.

\item Tourangeau, R., and Yan, T. (2007). 
Sensitive questions in surveys.
\emph{Psychological Bulletin}, 133: 859--883.

\item Van Buuren, S., Brand, J.P., Groothuis-Oudshoorn, C.G. and Rubin, D.B. (2006). 
Fully conditional specification in multivariate imputation.
\emph{Journal of statistical computation and simulation}, 76, 1049--1064.

\item Vercruyssen, A., Wuyts, C., and Loosveldt, G. (2017).
The effect of sociodemographic (mis)match between interviewers and respondents on unit and item nonresponse in Belgium.
\emph{Social Science Research}, 67: 229--238.

\item Von Hippel, P.T., (2020).
How many imputations do you need? A two-stage calculation using a quadratic rule. \emph{Sociological Methods \& Research}, 49, 699--718.

\item West, B.~T. and Blom, A.~G. (2017). 
Explaining interviewer effects: A research synthesis.
\emph{Journal of Survey Statistics and Methodology}, 5: 175--211.

\item Wuyts, C., and Loosveldt, G. (2017). 
The Interviewer in the respondent’s shoes: What can We learn from the way interviewers answer survey questions?.
\emph{Field Methods}, 29: 140--153.

\end{description}
\clearpage

%--------------------------------------------------------------------------------------------------
 \appendix 
%--------------------------------------------------------------------------------------------------
\section*{Appendix A: Tables}

%--------------------------------------------------------------------------------------------------

\begin{table}[!ht]
\caption{Estimates of the average marginal effects of the interviewer's expected response rate over the complete-case subsample by country through the CCA approach } 
\footnotesize
\begin{center}
\begin{tabular}{l|r@{\,}lr@{\,}lr@{\,}lr@{\,}l}
\hline
\multicolumn{1}{l|}{Country} &
\multicolumn{2}{c}{thinc2} &
\multicolumn{2}{c}{ypen1} &
\multicolumn{2}{c}{bacc} &
\multicolumn{2}{c}{home} \\
\hline
ES &         0.042&  &      0.092&**&      0.098&**&      0.069&* \\
   &       (0.030)&  &    (0.028)&  &    (0.034)&  &    (0.034)&  \\ 
IT &         0.101&**&      0.098&**&      0.262&**&      0.017&  \\
   &       (0.017)&  &    (0.016)&  &    (0.025)&  &    (0.021)&  \\ 
GR &         0.075&* &      0.107&**&     -0.085&  &     -0.043&  \\
   &       (0.035)&  &    (0.039)&  &    (0.049)&  &    (0.041)&  \\ 
PT &        -0.036&  &      0.020&  &     -0.033&  &      0.131&* \\
   &       (0.042)&  &    (0.034)&  &    (0.049)&  &    (0.055)&  \\ 
PL &         0.080&  &      0.146&**&      0.087&  &      0.010&  \\
   &       (0.051)&  &    (0.041)&  &    (0.064)&  &    (0.063)&  \\ 
SI &         0.139&**&      0.039&  &      0.177&**&      0.224&**\\
   &       (0.024)&  &    (0.021)&  &    (0.028)&  &    (0.028)&  \\ 
AT &         0.013&  &      0.046&**&      0.028&  &      0.044&  \\
   &       (0.018)&  &    (0.017)&  &    (0.022)&  &    (0.030)&  \\ 
DE &         0.040&**&      0.032&* &      0.096&**&      0.035&  \\
   &       (0.016)&  &    (0.014)&  &    (0.018)&  &    (0.019)&  \\ 
BE &         0.025&  &      0.046&**&      0.031&  &      0.030&  \\
   &       (0.015)&  &    (0.015)&  &    (0.020)&  &    (0.018)&  \\ 
LU &         0.082&* &      0.057&  &      0.174&**&     -0.005&  \\
   &       (0.039)&  &    (0.043)&  &    (0.041)&  &    (0.038)&  \\ 
SE &         0.004&  &     -0.015&  &      0.023&  &      0.037&* \\
   &       (0.022)&  &    (0.025)&  &    (0.027)&  &    (0.019)&  \\ 
EE &         0.042&  &     -0.053&**&      0.125&**&      0.116&**\\
   &       (0.023)&  &    (0.018)&  &    (0.031)&  &    (0.035)&  \\ 

\hline
\end{tabular}
\end{center}
\label{TAB:MDPATT_CC}
\begin{center}
\vspace{-8mm}
\parbox{105mm}{\footnotesize}
\end{center}
\end{table}
\clearpage

\begin{table}[!ht]
\caption{Estimates of the average marginal effects of the interviewer's expected response rate over the complete-case subsample by country through the FI\_MI approach} 
\footnotesize
\begin{center}
\begin{tabular}{l|r@{\,}lr@{\,}lr@{\,}lr@{\,}l}
\hline
\multicolumn{1}{l|}{Country} &
\multicolumn{2}{c}{thinc2} &
\multicolumn{2}{c}{ypen1} &
\multicolumn{2}{c}{bacc} &
\multicolumn{2}{c}{home} \\
\hline

ES &         0.079&  &      0.096&  &      0.093&  &      0.079&  \\
   &       (0.074)&  &    (0.072)&  &    (0.055)&  &    (0.068)&  \\ 
IT &         0.085&**&      0.073&**&      0.211&**&      0.024&  \\
   &       (0.026)&  &    (0.019)&  &    (0.042)&  &    (0.029)&  \\ 
GR &         0.038&  &      0.062&  &      0.003&  &     -0.102&* \\
   &       (0.054)&  &    (0.047)&  &    (0.062)&  &    (0.051)&  \\ 
PT &         0.080&  &      0.076&  &      0.038&  &      0.151&  \\
   &       (0.074)&  &    (0.041)&  &    (0.072)&  &    (0.077)&  \\ 
PL &         0.081&  &      0.095&  &      0.120&  &      0.059&  \\
   &       (0.073)&  &    (0.050)&  &    (0.092)&  &    (0.101)&  \\ 
SI &         0.161&**&      0.117&* &      0.191&**&      0.218&**\\
   &       (0.051)&  &    (0.056)&  &    (0.048)&  &    (0.059)&  \\ 
AT &         0.020&  &      0.042&* &      0.036&  &      0.035&  \\
   &       (0.025)&  &    (0.017)&  &    (0.028)&  &    (0.032)&  \\ 
DE &         0.038&* &      0.023&  &      0.085&**&      0.036&  \\
   &       (0.017)&  &    (0.015)&  &    (0.022)&  &    (0.022)&  \\ 
BE &         0.033&  &      0.059&**&      0.031&  &      0.051&* \\
   &       (0.018)&  &    (0.017)&  &    (0.025)&  &    (0.021)&  \\ 
LU &         0.088&* &      0.054&  &      0.116&  &      0.023&  \\
   &       (0.041)&  &    (0.041)&  &    (0.061)&  &    (0.046)&  \\ 
SE &        -0.003&  &     -0.011&  &      0.039&  &      0.036&  \\
   &       (0.025)&  &    (0.029)&  &    (0.032)&  &    (0.020)&  \\ 
EE &         0.037&  &     -0.019&  &      0.082&  &      0.046&  \\
   &       (0.037)&  &    (0.027)&  &    (0.066)&  &    (0.068)&  \\ 
\hline
\end{tabular}
\end{center}
\label{TAB:AME_ERR_FI_MI}
\begin{center}
\vspace{-8mm}
\parbox{105mm}{\footnotesize}
\end{center}
\end{table}
\clearpage

\begin{table}[!ht]
\caption{Estimates of the average marginal effects of interviewer's expected response rate over complete-case subsample by country through BBMA\_BIC approach} 
\footnotesize
\begin{center}
\begin{tabular}{l|r@{\,}lr@{\,}lr@{\,}lr@{\,}l}
\hline
\multicolumn{1}{l|}{Country} &
\multicolumn{2}{c}{thinc2} &
\multicolumn{2}{c}{ypen1} &
\multicolumn{2}{c}{bacc} &
\multicolumn{2}{c}{home} \\
\hline

ES &         0.079&  &      0.096&  &      0.093&  &      0.079&  \\
   &       (0.074)&  &    (0.072)&  &    (0.055)&  &    (0.068)&  \\ 
IT &         0.085&**&      0.073&**&      0.211&**&      0.024&  \\
   &       (0.026)&  &    (0.019)&  &    (0.042)&  &    (0.029)&  \\ 
GR &         0.038&  &      0.062&  &      0.003&  &     -0.102&* \\
   &       (0.054)&  &    (0.047)&  &    (0.062)&  &    (0.051)&  \\ 
PT &         0.080&  &      0.076&  &      0.038&  &      0.151&* \\
   &       (0.074)&  &    (0.041)&  &    (0.072)&  &    (0.077)&  \\ 
PL &         0.081&  &      0.113&**&      0.120&  &      0.059&  \\
   &       (0.073)&  &    (0.042)&  &    (0.091)&  &    (0.101)&  \\ 
SI &         0.164&**&      0.072&* &      0.196&**&      0.218&**\\
   &       (0.046)&  &    (0.036)&  &    (0.043)&  &    (0.059)&  \\ 
AT &         0.020&  &      0.042&* &      0.036&  &      0.035&  \\
   &       (0.025)&  &    (0.017)&  &    (0.028)&  &    (0.032)&  \\ 
DE &         0.038&* &      0.023&  &      0.085&**&      0.036&  \\
   &       (0.017)&  &    (0.015)&  &    (0.022)&  &    (0.022)&  \\ 
BE &         0.033&  &      0.059&**&      0.031&  &      0.051&* \\
   &       (0.018)&  &    (0.017)&  &    (0.025)&  &    (0.021)&  \\ 
LU &         0.088&* &      0.054&  &      0.116&  &      0.023&  \\
   &       (0.041)&  &    (0.041)&  &    (0.061)&  &    (0.046)&  \\ 
SE &        -0.003&  &     -0.011&  &      0.039&  &      0.037&  \\
   &       (0.025)&  &    (0.029)&  &    (0.032)&  &    (0.020)&  \\ 
EE &         0.037&  &     -0.019&  &      0.082&  &      0.050&  \\
   &       (0.037)&  &    (0.027)&  &    (0.066)&  &    (0.066)&  \\ 

\hline
\end{tabular}
\end{center}
\label{TAB:AME_ERR_BBMA_BIC}
\begin{center}
\vspace{-8mm}
\parbox{105mm}{\footnotesize}
\end{center}
\end{table}
\clearpage

\begin{table}[!ht]
\caption{Estimates of the average marginal effects of interviewer's expected response rate over complete-case subsample by country through BBMA\_AIC approach } 
\footnotesize
\begin{center}
\begin{tabular}{l|r@{\,}lr@{\,}lr@{\,}lr@{\,}l}
\hline
\multicolumn{1}{l|}{Country} &
\multicolumn{2}{c}{thinc2} &
\multicolumn{2}{c}{ypen1} &
\multicolumn{2}{c}{bacc} &
\multicolumn{2}{c}{home} \\
\hline

ES &         0.042&  &      0.092&**&      0.098&**&      0.069&* \\
   &       (0.030)&  &    (0.028)&  &    (0.034)&  &    (0.035)&  \\ 
IT &         0.101&**&      0.098&**&      0.262&**&      0.017&  \\
   &       (0.017)&  &    (0.016)&  &    (0.025)&  &    (0.021)&  \\ 
GR &         0.075&* &      0.107&**&     -0.085&  &     -0.043&  \\
   &       (0.035)&  &    (0.039)&  &    (0.049)&  &    (0.042)&  \\ 
PT &        -0.036&  &      0.020&  &     -0.033&  &      0.132&* \\
   &       (0.042)&  &    (0.034)&  &    (0.049)&  &    (0.056)&  \\ 
PL &         0.065&  &      0.145&**&      0.085&  &     -0.006&  \\
   &       (0.050)&  &    (0.041)&  &    (0.061)&  &    (0.065)&  \\ 
SI &         0.144&**&      0.039&  &      0.194&**&      0.222&**\\
   &       (0.024)&  &    (0.021)&  &    (0.029)&  &    (0.028)&  \\ 
AT &         0.014&  &      0.045&**&      0.026&  &      0.038&  \\
   &       (0.019)&  &    (0.017)&  &    (0.022)&  &    (0.031)&  \\ 
DE &         0.040&* &      0.029&* &      0.093&**&      0.036&  \\
   &       (0.016)&  &    (0.015)&  &    (0.018)&  &    (0.021)&  \\ 
BE &         0.025&  &      0.057&**&      0.035&  &      0.047&* \\
   &       (0.016)&  &    (0.017)&  &    (0.020)&  &    (0.020)&  \\ 
LU &         0.083&* &      0.067&  &      0.154&**&      0.016&  \\
   &       (0.039)&  &    (0.043)&  &    (0.040)&  &    (0.044)&  \\ 
SE &         0.002&  &     -0.017&  &      0.024&  &      0.038&* \\
   &       (0.023)&  &    (0.025)&  &    (0.028)&  &    (0.019)&  \\ 
EE &         0.040&  &     -0.050&* &      0.119&**&      0.112&**\\
   &       (0.025)&  &    (0.020)&  &    (0.032)&  &    (0.034)&  \\ 

\hline
\end{tabular}
\end{center}
\label{TAB:AME_ERR_BBMA_AIC}
\begin{center}
\vspace{-8mm}
\parbox{105mm}{\footnotesize 
}
\end{center}
\end{table}
\clearpage

%--------------------------------------------------------------------------------------------------
\section*{Appendix B: Figures}
%------------------------------------------------------------------------------------------------------
%--------------------------------------------------------------------------------------------------

\end{document}